\pdfoutput=1
\documentclass[12pt]{iopart}
\usepackage[utf8]{inputenc}
\usepackage{iopams}
\usepackage{amsfonts}
\usepackage{graphicx}
\usepackage{hyperref}
\usepackage[euler]{textgreek}

\newcommand{\muG}{\textmu Gal}

\usepackage{tikz}

\usepackage{paralist}

\usepackage{caption}
\begin{document}
\maketitle
\paper{On the effect of the light propagation within the corner-cube reflector of absolute gravimeters}
\author{V D Nagornyi}

\address{WSP USA Inc., One Penn Plaza, New York, NY 10019, USA}

\ead{vn2@member.ams.org}
\author{S Svitlov}
\address{Institut f\"ur Erdmessung, Leibniz Universit\"at Hannover, Schneiderberg 50, D-30167 Hanover, Germany} 
\ead{svitlov@ife.uni-hannover.de}
\begin{abstract}
We have assessed the implications of the in-cube light propagation effect in absolute gravimeters, and found it contradictory to existing theoretical and experimental data. We maintain that the `effect' is a bias of the quadratic coefficient that appears when the model of the trajectory ignores insensitivity of the laser interferometer to the constant part of the phase. The experimental bias obtained by fitting such an inadequate model was found in full agreement with the theoretical evaluations of the effect.
\end{abstract}
%
%
\pagestyle{empty}
%
%
%
%
\section{Introduction}
The light propagation within the corner cube reflector has recently been reported to significantly affect the acceleration measured by absolute gravimeters \cite{ashby2018}. The magnitude of the effect was found to be
\begin{equation}
\label{eq_the_effect}
\delta g =  \gamma (D n - d) = \gamma \, L,
\end{equation}
where $\gamma$ is the vertical gravity gradient, $D$ is the cube's depth from its face to the corner, $d$ is the distance from the face to the cube's centre of mass,
$n$ is the refractive index of the cube's material. Assuming $D$=1.75~cm, $\gamma$ = 307~E\footnote{1 E = $10^{-9}$ s$^{-1}$}, $n$=1.57, $d=D/4$ \cite{ashby2018}, the effect reaches the value of 7 \muG\footnote{1 \muG = $10^{-8}$ ms$^{-2}$}. Several authors have found problems with the reasoning substantiating the effect  \cite{kren2018, nagornyi2018, svitlov2018}, but the arguments of the critique were declined \cite{ashby2018a, ashby2018b, ashby2018c, ashby2018d}. The discussed issue directly affects the realisation of the new definition of the base SI unit of mass, the kilogram \cite{CGPM26-2018, vancamp2018}, therefore, the question of the existence of the effect needs to be resolved.

This paper considers the effect from different points of view. We first describe several conceptual tests pointing at contradictions between the effect and existing knowledge. We then look into the misconceptions that led to the discovery of the effect, including its alleged agreement with experimental data. Several times in previous discussions the method of `matching polynomials', used to evaluate the effect, came into focus. We examine the method in the Appendix.
\section{Reality tests for the effect}
\subsection{Test 1: Gravity surveys with different types of instruments}
\begin{figure}[ht]
\centering
\small
\input{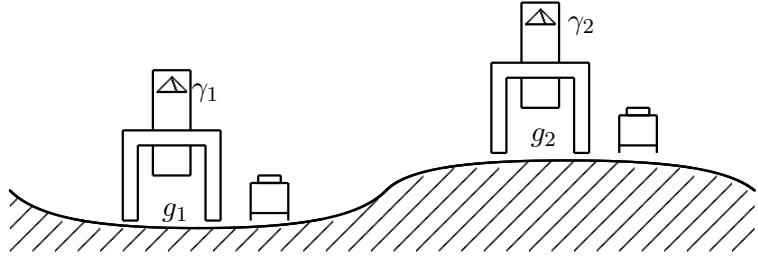}
  \caption[short title]
  {
  \quad\parbox[t]{12cm} {Gravity difference between two sites with different values of vertical gravity gradient. The difference measured by the corner-cube gravimeter should include the in-cube light propagation effect. 
  }
  }
\label{pic_test_grav}
\end{figure}

Let's consider a gravimetric link between two sites with absolute gravity values of $g_1$ and $g_2$,  and vertical gravity gradient values of $\gamma_1$ and $\gamma_2$ (fig.\ref{pic_test_grav}).
Assuming the effect (\ref{eq_the_effect}) exists, every measurement by a corner-cube gravimeter should include a systematic error of $\gamma  L$, not existent for other types of instruments. Because of that, the gravity difference measured by the corner-cube gravimeter (cc) will deviate from that measured by any other instrument (oth) by
\begin{equation}
\label{eq_grav_diff}
\left(g_2 -  g_1 \right)_\textrm{cc}
- \left( g_2 - g_1 \right)_\textrm{oth}
= L \, \left(\gamma_2 - \gamma_1 \right).
\end{equation}
Depending on various factors, the vertical gradients can differ by more than 250~E, even for the sites located within the same region \cite{marti2014}, so the difference (\ref{eq_grav_diff}) can reach 6 \muG. This value significantly exceeds the precision of modern gravity surveys and could not have been overlooked, if the effect (\ref{eq_the_effect}) existed.
\subsection{Test 2: Interaction of light and gravity}
The influence of gravity on light is known to be weak, only manifesting itself on such objects as galaxies, black holes, etc. The effect (\ref{eq_the_effect}) means that -- not even the Earth's gravity, but its vertical gradient, a signal several orders of magnitude weaker, affects the light propagation in an observable and measurable way. The existence of the effect would therefore contradict the current understanding of the light/gravity interaction \cite{gasperini2017}.
\subsection{Test 3: Position of the reflector's centre of mass}
The effect (\ref{eq_the_effect}) depends on the position of the reflector's centre of mass, located at a distance $d$ from it's face. For the experimental confirmation of the effect, the value of $d$ was assumed to be $D/4$, yielding the agreement with the theory within 0.001 \muG~ \cite{ashby2018}. However, in the gravimeters FG5, for which the experiment was conducted, the centre of mass of the falling object is adjusted to coincide with the reflector's optical centre \cite{niebauer2015}, located from the face at a distance $D/n$, rather than $D/4$ \cite{peck1948}. If the effect (\ref{eq_the_effect}) existed, this difference should have caused a major disagreement of about 1.8 \muG~between the  theory and the experiment. We further discuss the agreement between the theory and the experiment in Section \ref{sec_agreement}.
\subsection{Test 4: Polynomial properties of the phase of the reflected beam}
\label{sec_test_phase}
In absolute gravimeters, the information about the cube motion is deduced from the phase of the reflected laser beam. When the cube is at rest with respect to the beam source, the phase is not changing. The motion with a constant velocity creates a phase linearly increasing in time, i.e. the signal of a constant frequency. Similarly, an accelerated motion creates a phase proportional to the second degree of time, and the deviations from the uniform acceleration will cause fluctuations of the proportionality coefficient. The phase found in \cite{ashby2018} does not contain any previously unknown distortions of the quadratic coefficient \cite{nagornyi2018}, thus revealing no new effects.
\\
\\
We have looked into the in-cube propagation effect from different points of view and found it contradictory to existing theoretical and experimental knowledge. We now look into the misconceptions that led to the discovery of the effect.
\section{A constant phase is not measurable in displacement interferometry}
\label{sec_displ_intrfr}
The analysis leading to the effect starts from the phase of the reflected signal at the point of recombination \cite{ashby2018}:
\begin{equation}
\label{eq_the_phase}
\eqalign{
\phi(T,0)=\frac{2( D n-d) \Omega}{c}
+\frac{2 \Omega Z_0}{c}
-\frac{2 (D n-d) \Omega V_0}{c^2}
-\frac{2 \Omega Z_0 V_0}{c^2}\\
+T(-\Omega +\frac{2\Omega V_0}{c}-\frac{2\Omega V_0^2}{c^2}+\frac{2\Omega g ( D n-d)}{c^2}+\frac{2 g \Omega Z_0}{c^2})\\
+T^2(-\frac{g \Omega}{c}+\frac{3 g \Omega V_0}{c^2})-\frac{g^2 \Omega T^3}{c^2}\\
+\gamma \Omega \bigg( T(-\frac{2 ( D n-d)Z_0}{c^2}-\frac{2 Z_0^2}{c^2})
+T^2(\frac{Z_0}{c}-\frac{(D n-d) V_0}{c^2}-\frac{4V_0Z_0}{c^2})\\
+T^3(\frac{V_0}{3c}
-\frac{g( D n-d)}{ 3c^2}-\frac{4 V_0^2}{3 c^2}
+\frac{7g Z_0}{3c^2})
+T^4(-\frac{g}{12 c}+\frac{5 g V_0}{4 c^2})-\frac{g^2 T^5}{4 c^2}  \bigg).
}
\end{equation}
Here $\Omega T$ is the phase of the direct beam sent to the cube reflector, $Z_0, V_0, g$ are parameters of the trajectory at the moment $T=0$ : the distance from the cube to the point of recombination, the velocity, and the acceleration; $\gamma$ is the vertical gravity gradient, a known quantity.
The components of (\ref{eq_the_phase}) proportional to the $c^{-2}$ are caused by the delays in the light propagation. As established by \cite{nagornyi2018, svitlov2018}, these components produce the same effect as found earlier \cite{nagornyi2011}, so we drop them from further considerations.
By adding the phase $\Omega T$ of the reference beam, substituting $Dn-d \equiv L$, and expressing the phase $\Omega$ via the laser wavelength $\lambda$ as $\Omega=2\pi c/\lambda$, we get a theoretical phase of the beat signal at the point of recombination:
\begin{equation}
\label{eq_phase_beat}
\phi(T,0) = \frac{4\pi}{\lambda}
\left[
L + Z_0
+V_0 \,T
-\frac{g}{2}\,T^2
+\gamma \left( 
\frac{Z_0 T^2}{2} +\frac{V_0 T^3}{6}
-\frac{g T^4}{24} \right)
\right].
\end{equation}
As every period of the beat signal corresponds to the advancement of the reflector on the distance of $\lambda/2$, the path covered from the start of the measurement to the moment $T_i$ will be
\begin{eqnarray}
\label{eq_distance_at_Ti}    
Z(T_i) = 
\frac{\lambda}{4\pi}
\left[
\phi(T_i,0) - \phi(0,0)
\right] \nonumber \\
= 
V_0 \,T_i
-\frac{g}{2}\,T_i^2
+\gamma \left( 
\frac{Z_0 T_i^2}{2} +\frac{V_0 T_i^3}{6}
-\frac{g T_i^4}{24} \right).
\end{eqnarray}
The constant components $L+Z_0$ have been cancelled in the difference (\ref{eq_distance_at_Ti}), thus highlighting insensitivity of the displacement interferometry to a constant phase.
\section{The coordinates for data collection must originate at the trajectory}
\begin{figure}[ht]
\centering
\small
\input{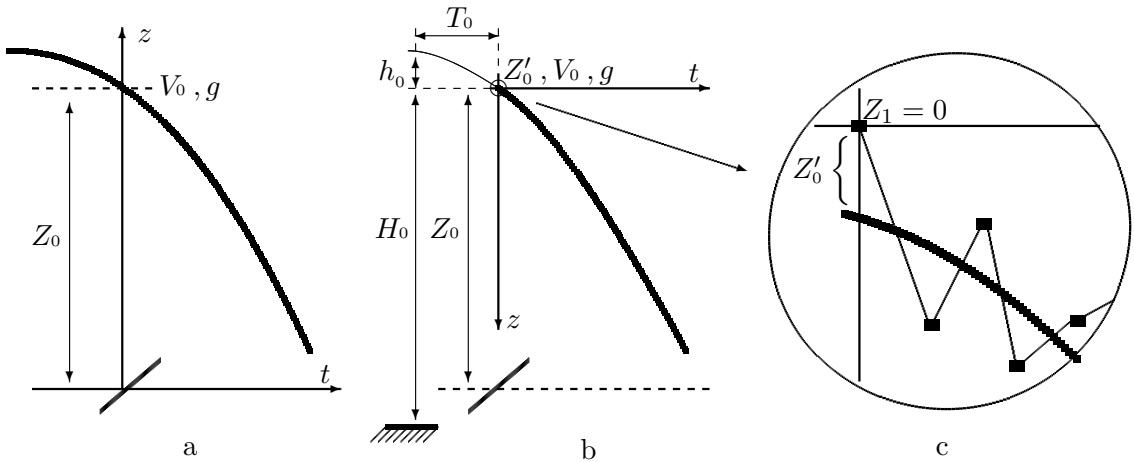}
  \caption
  {
  \quad\parbox[t]{12cm} {Coordinates used in absolute gravimetry:\\
  a -- coordinates for theoretical analysis can originate anywhere,\\
  b -- coordinates for actual measurements can only originate at the trajectory,\\
  c -- meaning of the initial displacement parameter in the coordinates used for the measurement.
  }
  }
\label{pic_2Z0}
\end{figure}

The phase (\ref{eq_the_phase}) was obtained in the coordinates originating at the beamsplitter, whereas the distance $Z_0$ from the beamsplitter to the first position coordinate (fig.\ref{pic_2Z0}a) was treated as an adjustable parameter of the model.
However, for collecting the data generated by the interferometer, the coordinates have to be moved to a new origin, because the distances (\ref{eq_distance_at_Ti}) can only be counted incrementally relative to a point located within the trajectory. The offset $Z_0$ linking the old and the new origins can only be established by an external ruler. In actual measurements the new origin is linked to the site reference mark, and the ruler is used to measure the distance $H_0+h_0$ from the mark to the resting position of the cube's optical centre (fig.\ref{pic_2Z0}b). Depending on how the time coordinate is counted, the distance $h_0$ from the resting position to the new origin can be computed as $h_0 = V_0^2/(2g)$ (if the first time coordinate is set to zero) \cite{rothleitner2009}, or as $h_0 = gT_0^2/2 - V_0T_0$ (if the first time coordinate is placed at the known initial offset $T_0$, fig.\ref{pic_2Z0}b) \cite{svitlov2018}.
To obtain the acceleration $g$, multiple positions $Z_i$ are measured at the moments $T_i$. A model of motion is then fitted to the data pairs $\{T_i, Z_i\}$ to minimise the errors $\varepsilon_i$. The model usually is 
\begin{equation}
\label{eq_model_LSS}    
Z_i = Z'_0 + V_0 \,T_i
+ \frac{g}{2}\,T_i^2
+\gamma \left( 
\frac{Z'_0 T_i^2}{2} +\frac{V_0 T_i^3}{6}
-\frac{g T_i^4}{24} \right) + \varepsilon_i.
\end{equation}
Though the interferometer is insensitive to a constant displacement, the model includes the parameter $Z'_0$, which, however, has nothing to do with the distances $Z_0$ or $Z_0+L$. The parameter $Z'_0$ is used due to the random noise in the coordinates $Z_i$ and serves to enhance statistical properties of the estimates by adding a degree of freedom to the model. The estimated value of $Z'_0$ itself is minuscule and of little practical importance, as it only catches the noise. More precisely, $Z'_0$ equals to the first residual $\varepsilon_1$ (fig.\ref{pic_2Z0}c), which is usually much less than 1 \textmu m. For that reason the model (\ref{eq_model_LSS}) is often simplified to
\begin{equation}
\label{eq_model_LSS1}    
Z_i = Z'_0 + V_0 \,T_i
+ \frac{g}{2}\,T_i^2
+\gamma \left( 
\frac{V_0 T_i^3}{6}
-\frac{g T_i^4}{24} \right) + \varepsilon_i.
\end{equation}
If the model (\ref{eq_model_LSS1}) is fitted to the data with the time coordinates shifted by the initial offset $T_0$, the parameter $g$ is automatically determined at the height $H_0+h_0$ (fig.\ref{pic_2Z0}b) \cite{svitlov2018}.

To sum up, the initial displacement parameter $Z’_0$  in the models (\ref{eq_model_LSS}) or (\ref{eq_model_LSS1}) cannot be used to estimate the values of $Z_0$, or $L$, or $Z_0+L$. None of the values can be found from the least-squares adjustment, if the distances are measured by means of the laser displacement interferometry. The problem with the analysis given in \cite{ashby2018} would become self-evident, if the paper reported on the  value of the parameter $Z_0$ found from the least-squares adjustment. The value must be too small to represent the distance between the beamsplitter and the cube. 
\section{The experiment confirms a bias due to inadequate model, not a real physical effect}
\label{sec_agreement}
As seen above, if the quantity $L$ is added to the right-hand side of (\ref{eq_model_LSS}), the resulting expression
\begin{equation}
\label{eq_model_LSS_L}    
Z_i = L + Z'_0 + V_0 \,T_i
+ \frac{g}{2}\,T_i^2
+\gamma \left( 
\frac{Z'_0 \,T_i^2}{2} +\frac{V_0 \, T_i^3}{6}
-\frac{g\, T_i^4}{24} \right) + \varepsilon_i
\end{equation}
could no longer adequately model the distances measured by a laser interferometer. If still fit to the trajectory data, the model would produce a biased estimates of the acceleration parameter. To find the bias, let's notice that the model (\ref{eq_model_LSS_L}) is equivalent to fitting the model (\ref{eq_model_LSS}) to the data $Z_i-L$:
\begin{equation}
\label{eq_model_LSS_L_}    
Z_i -L =  Z'_0 + V_0 \,T_i
+ \frac{g}{2}\,T_i^2
+\gamma \left( 
\frac{Z'_0\, T_i^2}{2} +\frac{V_0 \, T_i^3}{6}
-\frac{g\, T_i^4}{24} \right) + \varepsilon_i,
\end{equation}
i.e. to lowering the entire trajectory on $L$ units and, consequently, to shifting the estimated gravity on $\gamma L$ units. The same estimate of the effect was found in \cite{ashby2018} by the method of matching polynomials, which we consider in the Appendix. Processing of the real trajectories measured with the FG5 gravimeter \cite{ashby2018} has revealed the same shift, leading to the conclusion about the agreement between the theory and the experiment. It must be stressed, however, that the agreement was reached not between the theory and a real physical effect, but rather between the numerical and analytic verifications of the bias caused by the same inadequate model. 
\section{Conclusions}
We found the effect of light propagation within the corner-cube reflector of absolute gravimeters, introduced in \cite{ashby2018}, to be contradictory to existing theoretical and experimental data. The effect is caused by a faulty assumption that an absolute position of the reflector with respect to the beamsplitter, not only the position change, can be measured by the interferometer. The shift of the measured gravity caused by fitting the inadequate model was found in agreement with theoretical evaluations of the bias.

\appendix
\section{Matching polynomials}
The bias of the acceleration parameter estimated in section \ref{sec_agreement} by shifting the trajectory was confirmed in \cite{ashby2018}  by the method of `matching polynomials', not previously used in absolute gravimetry. Because it's beneficial to have more techniques for error analysis, we analyse the method here.
First, we recapture the essential steps of the technique, then apply it to evaluate the effect of dropping the vertical gravity gradient from the model.
\subsection{In-cube light propagation estimated with matching polynomials}
\label{sec_matching}
Let's consider a continuous trajectory corresponding to the phase model (\ref{eq_phase_beat}):
\begin{equation}
\label{eq_model_LSS2_cont}    
\varphi(Z_0, V_0, g, T, L)  = L + Z_0 + V_0 \,T
- \frac{g}{2}\,T^2
+\gamma \left( 
\frac{Z_0\, T^2}{2}
+\frac{V_0\, T^3}{6}
-\frac{g\, T^4}{24} \right) .
\end{equation}
Assuming that the term $L$ changes the estimates of $Z_0, V_0, g$ by $\delta Z_0, \delta V_0, \delta g$, the difference between (\ref{eq_model_LSS2_cont}) and the original trajectory $\varphi(Z_0, V_0, g, T, 0)$ will be
\begin{eqnarray}
\label{eq_Delta}    
\Delta = 
\varphi(Z_0 + \delta Z_0, V_0 + \delta V_0, g + \delta g, T, L) - 
\varphi(Z_0, V_0, g, T, 0) \\
=
L + \delta Z_0
+ \delta V_0 \,T
-\frac{\delta g}{2}\,T^2
+\gamma \left( 
\frac{\delta Z_0\,T^2}{2} +\frac{\delta V_0\, T^3}{6}
-\frac{\delta g\,T^4}{24} \right).
\end{eqnarray}
The difference, a 4-th degree polynomial in $T$, is minimised by requesting that all its coefficients would turn to zeroes \cite{ashby2018, ashby2018c}:
\begin{eqnarray}
T^0: \; L + \delta Z_0 = 0,\\
T^1: \; \delta V_0 = 0,\\
T^2: \; \frac{\delta g}{2} - \gamma\,\frac{\delta Z_0}{2} = 0,\\
T^3: \; \gamma\,\frac{\delta V_0}{6} = 0,\\
T^4: \; - \gamma\,\frac{\delta g}{24} = 0.
\end{eqnarray}
The above can be viewed as system of linear equations with variables $\delta Z_0,\,\delta V_0,\,\delta g$. The matrix form of the system is
\begin{equation}
\label{eq_A1}
\left(
\begin{array}{ccc}
1 & 0 & 0  \\
0 & 1 & 0  \\
-\frac{\gamma}{2} & 0 & \frac12  \\
0 & \frac{\gamma}{6} & 0  \\
0 & 0 & -\frac{\gamma}{24}   
\end{array}
\right)
\left(
\begin{array}{c}
\delta Z_0 \\
\delta V_0 \\
\delta g
\end{array}
\right)
-
\left(
\begin{array}{c}
-L \\
0 \\
0 \\
0 \\
0 
\end{array}
\right)
= 0,
\end{equation}
which matches the formula (10) of \cite{ashby2018c}\footnote{As we do not consider signal delays, the matrix $B$ of \cite{ashby2018c} in our case turns to zero.}. This is an over-defined system, and solutions for $\delta Z_0, \delta V_0, \delta g$ cannot be found to satisfy all 5 equations. However, we can find exact solutions for the first 3 equations:
\begin{eqnarray}
\delta Z_0 = -L,\\
V_0 = 0,\\
\delta g = -\gamma \, L, \label{eq_the_corr}
\end{eqnarray}
Substituting the above solutions into the original system (\ref{eq_A1}), the first four equations are solved exactly delivering (\ref{eq_the_effect}), while the fifth one is solved approximately, with the remaining term of the order of $\gamma^2$.
\subsection{Matching polynomials applied to the effect of vertical gravity gradient}
Following the approach described in \cite{ashby2018, ashby2018c}, we estimated in (\ref{eq_the_corr}) the adjustment to the quadratic coefficient best compensating the quantity $L$ added to the model.
We now try to use the same approach to find adjustment for another quantity --- the vertical gravity gradient. This adjustment was studied thoroughly by different methods (e.g. \cite{niebauer1989, nagornyi1995, timmen2003}). The results in good agreement have shown that if a model without the gradient is fitted to the trajectory, the bias of the estimated acceleration parameter would be $\delta g = \gamma h$, where $h$ is the effective measurement height of the gravimeter, roughly about one-third of the trajectory length.
The model with the gradient is
\begin{equation}
\label{eq_model_grad}    
\varphi(Z_0, V_0, g, T, \gamma)  = Z_0 + V_0 \,T
- \frac{g}{2}\,T^2
+\gamma \left( 
\frac{Z_0'T^2}{2}
+\frac{V_0 T^3}{6}
-\frac{g T^4}{24} \right) .
\end{equation}
Like before, we assume that omitting the gradient changes the coefficients $Z_0, V_0, g$ by $\delta Z_0, \delta V_0, \delta g$, and find the difference between two approximations
\begin{eqnarray}
\label{eq_Delta_of_gamma}    
\hspace*{-2cm}
\Delta = 
\varphi(Z_0 + \delta Z_0, V_0 + \delta V_0, g + \delta g, T, \gamma) - 
\varphi(Z_0, V_0, g, T, 0) \\
\hspace*{-1.5cm} =
\delta Z_0 + \delta V_0 \,T
-\frac{\delta g}{2}\,T^2
+\gamma \left( 
\frac{(Z_0+\delta Z_0) T^2}{2} +\frac{(V_0 + \delta V_0) T^3}{6}
-\frac{(g+\delta g) T^4}{24} \right)
\end{eqnarray}
Following the method, we request the coefficients at all powers to vanish:
\begin{eqnarray}
T^0: \; \delta Z_0 = 0\\
T^1: \; \delta V_0 = 0\\
T^2: \; \frac{\delta g}{2} - \gamma\,\frac{Z_0 + \delta Z_0}{2} = 0 \\
T^3: \; \gamma\,\frac{V_0 + \delta V_0}{6} = 0 \\
T^4: \; -\gamma\,\frac{g + \delta g}{24} = 0 \end{eqnarray}
The matrix notation similar to (\ref{eq_A1}) is

\begin{equation}
\label{eq_A2}
\left(
\begin{array}{ccc}
1 & 0 & 0  \\
0 & 1 & 0  \\
-\frac{\gamma}{2} & 0 & \frac12  \\
0 & \frac{\gamma}{6} & 0  \\
0 & 0 & -\frac{\gamma}{24}   
\end{array}
\right)
\left(
\begin{array}{c}
\delta Z_0 \\
\delta V_0 \\
\delta g
\end{array}
\right)
-
\left(
\begin{array}{c}
0 \\
0 \\
\gamma \frac{Z_0}{2} \\
- \gamma \frac{V_0}{6} \\
\gamma \frac{g}{24}
\end{array}
\right)
= 0
\end{equation}
The first three equations lead to the following solution:
\begin{eqnarray}
\delta Z_0 = 0\\
\delta V_0 = 0\\
\delta g = \gamma\, Z_0 = 0.
\end{eqnarray}
Because the result $\delta g = 0$, compared to the expected one $\delta g = \gamma h$, is obviously incorrect, we must conclude that `matching polynomials' cannot serve as viable method of analysis in absolute gravimetry.
%
%
%
\section*{References}
\end{document}